\documentclass[%
 reprint,
 amsmath,amssymb,
 aps,
]{revtex4-2}

\usepackage{graphicx}
\usepackage{dcolumn}
\usepackage{bm}
\usepackage{hyperref}
\begin{document}

\title{Mechanism underlying the scaling law of home-return probability in human mobility}

\author{Haoying Niu}
\author{Xiao-Yong Yan}
 \email{yanxy@bjtu.edu.cn}
\affiliation{School of Systems Science, Beijing Jiaotong University, Beijing, China 
}%

\begin{abstract}

Individual daily mobility exhibits a striking scaling law: the probability of returning home after a tour of $l$ locations decays as $P_{\rm ret}(l)\sim l^{-\gamma}$. While the tour-terminate-continue (TTC) model reproduces this behavior, it relies on this power law as an empirical input, leaving the microscopic origin of $\gamma$ unresolved. Here we show that this scaling emerges from a utility trade-off governed by cognitive constraints. By invoking the principle of least effort, we demonstrate that individual activity priorities follow Zipf's law, $p(r)\sim r^{-\nu}$, which directly dictates the sublinear accumulation of tour utility, $U_L(l)\sim l^{1-\nu}$. Luce's choice rule then yields \(P_{\rm ret}(l)\sim l^{-(1-\nu)}\), giving the exact exponent $\gamma = 1 - \nu$. Agent-based simulations confirm this analytical relation. Our framework bridges the gap between individual cognitive constraints and the scaling law of tour behavior, providing a microscopic theoretical underpinning for human mobility.

\end{abstract}

\maketitle


Understanding human mobility is a core research topic across urban science, transportation engineering, and complex systems \cite{mata,song,barth,mob}. A central concept in capturing daily mobility is the individual tour, defined as the sequence of trips originating from home (a fixed base location), visiting several locations, and eventually returning home.
Recently, massive trajectory data has revealed statistical regularities in individual tour behavior \cite{lin}, among which the scaling law of tour length is the most striking. 

To explain this phenomenon, Lin et al proposed a tour-terminate-continue (TTC) model \cite{lin}, in which at each step $l$, an individualmakes a binary decision: either terminate the tour and return home, or continue tour to the next location.
The core of the TTC model relies on the empirical specification of the home-return probability, $P_{\mathrm{ret}}$. Empirical evidence indicates that the home-return probability decays as a power law with the tour length $l$:
\begin{equation}\label{eq1}
P_{\mathrm{ret}}(l) = \rho l^{-\gamma},
\end{equation}
where $\rho$ denotes a prefactor and $\gamma \in (0, 1)$ is the scaling exponent.

Although the TTC model successfully reproduces the distribution of tour lengths using Eq. \eqref{eq1}, it treats the power-law decay of $P_{\mathrm{ret}}(l)$ as an exogenous empirical input. A key open question therefore arises: why does the home-return probability follow a power law, and what determines the exponent $\gamma$? Without answering this, the TTC model remains a phenomenological description rather than a mechanistic explanation.

In this work, we argue that individual TTC decisions are governed by a trade-off between two competing utilities. Let $U_H$ be the utility of returning home, a constant term representing rest and sense of security. Let $U_L(l)$ denote the cumulative marginal utility derived from the external tour up to step $l$.
Under the assumption that the individual compares the relative attraction of returning home versus continuing tour, a simpler Luce choice rule \cite{luce} gives
\begin{equation}\label{eq2}
	P_{\mathrm{ret}}(l) = \frac{U_H}{U_H + U_L(l)}
\end{equation}
when utilities are treated as positive “weights”. This parsimonious form has been widely used in binary decision models of cognitive science.

The functional form of $P_{\mathrm{ret}}(l)$ is entirely determined by the growth pattern of $U_L(l)$. If external utilities were homogeneous, $U_L(l)$ would grow linearly with $l$, leading to $P_{\mathrm{ret}} \sim l^{-1}$. However, the empirical observation of $\gamma < 1$ \cite{lin} dictates that $U_L(l)$ must exhibit sublinear growth. This sublinearity implies a diminishing marginal utility of exploration: as the tour proceeds, the additional utility gained from visiting the next location decreases. The question then reduces to: what is the microscopic origin of this specific sublinear accumulation of utility?

We attribute the sublinear utility growth to the heterogeneous distribution of individual activity priorities. Individuals retrieve activity purposes from their cognitive system, which we model as an optimal coding process following Mandelbrot's information-theoretic framework \cite{mand}.
To minimize average cognitive effort (the principle of least effort \cite{zipf}), the brain encodes frequent purposes with short retrieval paths and rare purposes with long paths. 
Under the constraint of instantaneous decodability (optimal prefix code, e.g., Huffman coding \cite{huf}), Shannon's source coding theorem \cite{shan} dictates that the minimal retrieval cost (code length) for a purpose with probability $p(r)$ satisfies $C(r) \geqslant -\log p(r)$. For optimal prefix codes satisfying the Kraft–McMillan inequality \cite{mcm}, this bound is asymptotically tight: $C(r) \approx -\log p(r)$. Consequently, sorting purposes by priority rank $r$ (where $r=1$ is highest priority) allocates shorter codes to more probable events, yielding the logarithmic scaling
\begin{equation}\label{eq3}
	C(r) \propto \ln r.
\end{equation}

Minimizing the expected cognitive cost under an entropy constraint yields the optimal retrieval probability, reflecting inherent priority weights:
\begin{equation}\label{eq4}
	p(r) \propto \mathrm{e}^{-C(r)} = r^{-\nu}.
\end{equation}
Accordingly, Zipf’s law emerges at the individual level as a natural analytical outcome of cognitive least effort, where $\nu$ is the cognitive heterogeneity exponent.

Crucially, the priority rank $r$ derived from cognitive least effort is not merely an abstract list of necessities; it is an integrated evaluation that incorporates spatial accessibility during the cognitive encoding phase. From the perspective of bounded rationality, spatially proximate destinations require less cognitive effort to retrieve and execute. Therefore, the brain naturally assigns them higher effective priority within the Zipfian distribution. This cognitive pre-weighting implies that the sequence of tour steps $l$ naturally aligns with the integrated priority $r$ (i.e., $l = r$). When individuals choose ``the most convenient way to walk", they are executing the sequence dictated by this cognitively optimized ranking. In this framework, spatial convenience does not conflict with priority ranking; rather, it serves as the cognitive optimization mechanism that shapes the observed sequence of visits.

Since the marginal utility of visiting a purpose is proportional to this integrated priority, we have $u(r) \propto p(r) \propto r^{-\nu}$. The cumulative marginal utility after $l$ steps is the sum over the top $l$ ranks:
\begin{equation}\label{eq5}
	U_L(l) = \sum_{r=1}^{l} u(r) \propto \sum_{r=1}^{l} r^{-\nu}.
\end{equation}

To evaluate the asymptotic behavior for large $l$, we distinguish between three regimes of the Zipf exponent $\nu$:

Case 1: $\nu > 1$. The sum converges to the Riemann zeta function $\zeta(\nu)$. As $l \to \infty$, $U_L(l) \sim \zeta(\nu)$, a constant. This would imply $P_{\mathrm{ret}}$ approaches a constant, contradicting the empirical power-law decay.

Case 2: $\nu = 1$. The harmonic sum yields $U_L(l) \sim \ln l$ for large $l$. Substituting this logarithmic growth into Eq. \eqref{eq2} gives $P_{\mathrm{ret}(l)} \sim 1/\ln l$, which decays slower than power law and does not converge to a constant. 

Case 3: $0 < \nu < 1$. The sum diverges as $l$ increases. Integral approximation yields
\begin{equation}\label{eq6}
	U_L(l) \approx K \int_{1}^{l} r^{-\nu} \mathrm{d}r = \frac{K}{1-\nu} \left( l^{1-\nu} - 1 \right) \sim \frac{Kl^{1-\nu}}{1-\nu},
\end{equation}
where $K$ is the proportionality constant relating marginal utility to priority rank (i.e., $u(r) = Kr^{-\nu}$).
This sublinear growth is the mathematical manifestation of the diminishing marginal utility.

Substituting Eq. \eqref{eq6} into Eq. \eqref{eq2} gives
\begin{equation}\label{eq7}
	P_{\mathrm{ret}}(l) \approx \frac{U_H}{U_H + K l^{1-\nu}/(1-\nu)}.
\end{equation}
 In the asymptotic regime where the tour length is large (i.e., $K l^{1-\nu}/(1-\nu) \gg U_H$), taking the leading-order term yields the scaling relation
\begin{equation}\label{eq8}
	P_{\mathrm{ret}}(l) \sim  l^{-(1-\nu)}.
\end{equation}
Comparing Eq. \eqref{eq8} with Eq. \eqref{eq1}, we arrive at the scaling exponent
\begin{equation}\label{eq9}
	\gamma = 1 - \nu.
\end{equation}
Given that $\nu \in (0,1)$, it naturally follows that $\gamma \in (0,1)$, which is fully consistent with empirical observations. 
We thus conclude that the power-law decay of home-return probability arises from sublinear accumulation of utility governed by the individual cognitive Zipf's law.

\begin{figure*}[htbp]
	\centering
	\includegraphics[width=1\textwidth]{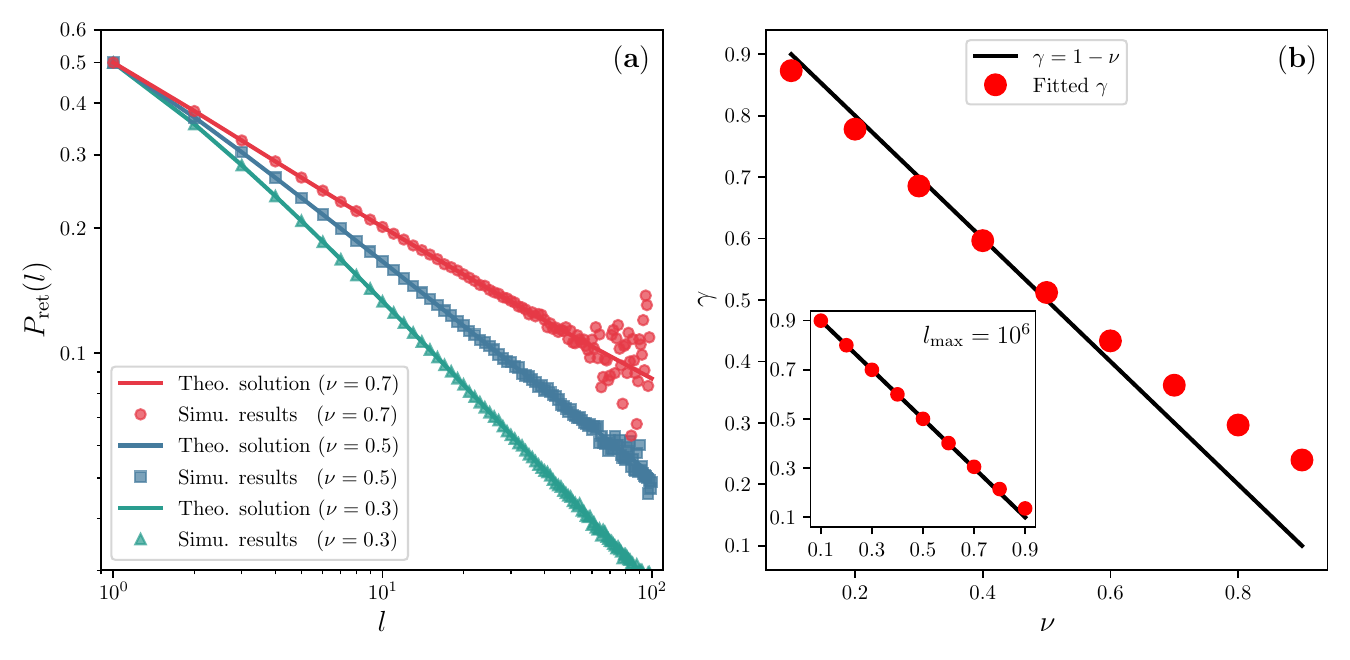}
	\caption{
		Simulation results for the home-return probability and its corresponding scaling exponent in the TTC decision-making process. (a) Comparison of the home-return probability $P_{\mathrm{ret}}(l)$ between theoretical solutions (lines) and simulation results (markers) for various $\nu$.  (b) Fitted scaling exponent $\gamma$ versus $\nu$. The line represents the theoretical prediction $\gamma = 1 - \nu$.}
	\label{fig1} 
\end{figure*}

To further validate the aforementioned theoretical results, we conduct agent-based simulations of the TTC decision-making process for individuals. Specifically, the number of agents is set to $10^8$, and the maximum tour length is set to $l_{\mathrm{max}} = 100$. For simplicity, the utility of home is set to $U_H = 1$, and the utility of the $l$-th visited node is set to $l^{-\nu}$, corresponding to the coefficient $K=1$. Figure \ref{fig1}(a) compares simulated $P_{\mathrm{ret}}(l)$ with theoretical solutions calculated via the discrete summation $U_L(l) =  \sum_{r=1}^{l} r^{-\nu}$. Simulations agree well with theory for small $l$, while noticeable fluctuations emerge at large $l$, particularly for higher $\nu$. This is attributed to the extremely small number of individuals who remain engaged in the TTC decision-making process when the tour length is very long, leading to increased statistical noise. Figure \ref{fig1}(b) displays the fitted scaling exponent $\gamma$ of the home-return probability $P_{\mathrm{ret}}(l)$ from the simulation results under different values of $\nu$. It can be seen that the simulation results generally agree with the theoretical predictions (Eq. \eqref{eq9}), although slight deviations still exist due to the finite $l_{\mathrm{max}}$. As $l_{\mathrm{max}}$ increases, $\gamma$ becomes closer to $1 - \nu$, which is clearly demonstrated by the inset in Fig. \ref{fig1}(b).

In summary, we have revealed the microscopic mechanism behind the scaling law of home-return probability in individual tour behavior. By bridging the TTC model with the principle of least effort at the individual cognitive level, we demonstrate that the power-law decay $P_{\mathrm{ret}}(l) \sim l^{-\gamma}$ emerges from the trade-off between fixed utility of returning home and the sublinearly growing marginal utility of continuing tour. Our key contribution is proving that this sublinear growth is governed by the individual's internal Zipf distribution of activity priorities, which itself is the analytical solution to optimal cognitive coding. This allows us to theoretically derive the scaling exponent as $\gamma = 1 - \nu$, providing a microscopic theoretical foundation for the empirical TTC model.

Although our theoretical framework well reproduces the macroscopic scaling pattern of human tour and return behavior, several limitations remain to be explored. 
First, while our model derives the scaling exponent from cognitive cost principles, it implicitly internalizes spatial friction through the concept of integrated priority. Rather than treating spatial distance as an external constraint that disrupts the tour, we argue that spatial accessibility modulates the cognitive cost of retrieving an activity purpose. Highly accessible locations are encoded with lower retrieval costs, effectively boosting their rank $r$ in the individual's preference hierarchy. Future work could explicitly quantify how spatial decay functions (e.g., distance decay) couple with cognitive heterogeneity $\nu$ to shape the observed mobility patterns, thereby unifying cognitive search theories with spatial interaction models \cite{mob}. Second, we focus primarily on individual-level optimal decision-making under the least effort principle, without incorporating social interactions or collective mobility patterns that could modulate empirical scaling exponents. Future research can extend the present microscopic derivation framework by introducing dynamic cognitive updating mechanisms and heterogeneous individual preferences, so as to generalize the theoretical scaling relation across diverse travel scenarios. Moreover, linking  our framework to reinforcement learning and bounded rationality may reveal whether this Zipfian priority ranking constitutes an optimal adaptive strategy in uncertain environments. Such efforts will strengthen the links between cognitive science, complex systems and urban research.

\begin{acknowledgments}
This work was supported by the National Natural Science Foundation of China (Grant Nos. 72271019, 71822102).
\end{acknowledgments}


\begin{thebibliography}{99}
\bibitem{mata}
M. C. Gonz\'{a}lez, C. A. Hidalgo, and A. L. Barab\'{a}si, Understanding individual human mobility patterns, Nature (London) 453, 779 (2008).
\bibitem{song}
C. Song, T. Koren, P. Wang, and A. L. Barab\'{a}si, Modelling the scaling properties of human mobility, Nat. Phys. 6, 818 (2010).
\bibitem{barth} 
M. Barth\'{e}lemy, Spatial networks,
Phys. Rep. 499, 1 (2010).
\bibitem{mob} 
H. Barbosa, M. Bath\'{e}lemy, G. Ghoshal, C. R. James, M. Lenormand, T. Louail, R. Menezes, J. J. Ramasco, F. Simini, and M. Tomasini, Human mobility: Models and applications, Phys. Rep. 734, 1 (2018).
\bibitem{lin} 
 X.-J. Lin, Y. Yang, W.-P. Nie, X.-Y. Yan, Scaling law of individual urban tour behavior, Phys. Rev. E 113, 034303 (2026).
\bibitem{luce} 
 R. D. Luce, Individual Choice Behavior: A Theoretical Analysis (Wiley, New York, 1959).
\bibitem{mand} 
B. Mandelbrot, An informational theory of the statistical structure of language, in W. Jackson, Communication Theory (Butterworths, London, 1953). 
\bibitem{zipf} 
G. K. Zipf, Human Behavior and the Principle of Least Ef-fort: An Introduction to Human Ecology (Addison-Wesley, Cambridge, MA, 2016).
\bibitem{huf} 
D. A. Huffman, A method for the construction of minimum-redundancy codes,  Proceedings of the IRE 40(9), 1098 (1952).
\bibitem{shan}
C. E. Shannon, A Mathematical Theory of Communication, Bell Syst. Tech. J. 27, 379 (1948).
\bibitem{mcm}
B. McMillan, Two inequalities implied by unique decipherability, IRE Trans. Inf. Theory 2(4), 115 (1956).
\end{thebibliography}
\end{document}